\documentclass[oldversion]{aa}
\usepackage{graphicx}
\def\H0 {$H_{\rm o}$}

\def\ffas {\hbox{$\,.\!\!^{\prime\prime}$}}

\def\ffs {\hbox{$\,.\!\!\!^{\rm s}$}}

\def\CH3C2H {\hbox{${\rm CH}_3{\rm C}_2{\rm H}$}} 

\def\ffas {\hbox{$\,.\!\!^{\prime\prime}$}}

\def \ga{\mathrel{\mathchoice   {\vcenter{\offinterlineskip\halign{\hfil
$\displaystyle##$\hfil\cr>\cr\sim\cr}}}
{\vcenter{\offinterlineskip\halign{\hfil$\textstyle##$\hfil\cr
>\cr\sim\cr}}}
{\vcenter{\offinterlineskip\halign{\hfil$\scriptstyle##$\hfil\cr
>\cr\sim\cr}}}
{\vcenter{\offinterlineskip\halign{\hfil$\scriptscriptstyle##$\hfil\cr
>\cr\sim\cr}}}}}

\begin{document}

\title{Ammonia in the hot core W\,51-IRS2:
       11 new maser lines and a maser component with a velocity drift}

\author{C. Henkel\inst{1,2} 
        \and 
        T.L. Wilson\inst{3} 
        \and 
        H. Asiri\inst{2}
        \and 
        R. Mauersberger\inst{4}}

\offprints{C. Henkel, \email{chenkel@mpifr-bonn.mpg.de}}

\institute{
  Max-Planck-Institut f{\"u}r Radioastronomie, Auf dem H{\"u}gel 69, 
  D-53121 Bonn, Germany
 \and 
  Astronomy Department, Faculty of Science, King Abdulaziz University, P.O. Box 80203, Jeddah, 
  Saudi Arabia
 \and
  Naval Research Laboratory, Code 7210, Washington, DC 20375, USA
 \and
  Joint ALMA Observatory, Avda. Alonso de C{\'o}rdova 3107, 
  Vitacura, Santiago de Chile, Chile
}
 
\date{Received date ; accepted date}
 
\abstract
{With the 100-m telescope at Effelsberg, 19 ammonia (NH$_3$) maser 
lines have been detected toward the prominent massive star forming region 
W51-IRS2. Eleven of these inversion lines, the ($J$,$K$) = (6,2), (5,3), 
(7,4), (8,5), (7,6), (7,7), (9,7), (10,7), (9,9), (10,9), and 
(12,12) transitions, are classified as masers for the first time in 
outer space. All detected masers are related to highly excited 
inversion doublets. The (5,4) maser originates from an inversion doublet
$\sim$340\,K above the ground state, while the (12,12) transition,
at $\sim$1450\,K, is the most highly excited NH$_3$ maser line so far 
known. Strong variability is seen not only in ortho- but also in 
para-NH$_3$ transitions. Bright narrow emission features are observed, 
for the first time, in (mostly) ortho-ammonia transitions, at 
$V_{\rm LSR}$ $\sim$ 45\,km\,s$^{-1}$, well separated from the 
quasi-thermal emission near 60\,km\,s$^{-1}$. These features were 
absent $\sim$25 years ago and show a velocity drift of about
+0.2\,km\,s$^{-1}$\,yr$^{-1}$. The component is likely related to 
the SiO maser source in W51-IRS2 and a possible scenario explaining 
the velocity drift is outlined. The 57\,km\,s$^{-1}$ component of 
the (9,6) maser line is found to be strongly linearly polarised. 
Maser emission in the ($J$,$K$) to ($J$+1,$K$) inversion 
doublets is strictly forbidden by selection rules for electric 
dipole transitions in the ground vibrational state. However, 
such pairs (and even triplets with ($J$+2,$K$)) are common 
toward W51-IRS2. Similarities in line widths and velocities 
indicate that such groups of maser lines arise from the same 
regions, which can be explained by pumping through vibrational 
excitation. The large number of NH$_3$ maser lines in W51-IRS2 is 
most likely related to the exceptionally high kinetic temperature 
and NH$_3$ column density of this young massive star forming region.}


\titlerunning{New ammonia masers}

\authorrunning{Henkel, C., Wilson, T.L., Asiri, H., Mauersberger, R.}

\maketitle

\section{Introduction}

Ammonia (NH$_3$) provides the unique opportunity to trace molecular cloud
excitation up to temperatures of $\sim $2000\,K by observing its
characteristic inversion transitions within a very limited frequency 
interval (20 -- 35\,GHz; e.g. Ho \& Townes 1983; Wilson et al. 2006,
2008). The inversion doublets arise from oscillations of the 
nitrogen nucleus through the plane of the three hydrogen nuclei. 
The frequencies of the lines connecting the two states of an
inversion-doublet depend on the total angular momentum $J$ and its 
projection on the molecular axis, $K$, with $K$ = 0,3,6,9... belonging 
to ortho-NH$_3$ and $K$ = 1,2,4,5,7...  representing para-NH$_3$. 
Thus in the vibrational ground state, inversion doublets are characterized 
by the values $J$ and $K$. The parities of the split levels differ to
allow for the dipole transitions ($J$,$K$) $\rightarrow$ ($J$,$K$)
(hereafter ($J$,$K$)), which are subject of this paper. 

Dozens of inversion lines can be detected, provided kinetic temperatures 
and ammonia column densities are high enough. These conditions prevail
in ``hot cores'', dense molecular clumps near sites of very recent massive 
star formation (e.g., Mauersberger et al. 1986a, 1988b; Henkel et al 1987b; 
Hermsen et al. 1988; Cesaroni et al. 1992; H{\"u}ttemeister et al. 1993, 
1995; Wilson et al. 1993, 2000; Zhang \& Ho 1997; Goddi et 
al. 2011). The NH$_3$ abundances are believed to be caused 
by dust grain mantle evaporation (e.g., Henkel et al. 1987a; Walmsley et al. 
1987; Brown et al. 1988). The warm dense clumps are characterized by 
temperatures $T_{\rm kin}$ $>$ 100\,K, $X$(NH$_3$) = $N$(NH$_3$)/$N$(H$_2$) 
$\sim$ 10$^{-5...-6}$ and source averaged ammonia column densities in 
excess of 10$^{18}$\,cm$^{-2}$.

Most but not all the ammonia inversion lines are thermally excited. ($J,K$) 
= (3,3) maser emission was first detected by Wilson et al. (1982) toward 
the star forming region W\,33. To date, several NH$_3$ maser lines have been
observed (e.g., Guilloteau et al. 1983; Madden et al. 1986; Mauersberger et 
al. 1988b; Hofner et al. 1994; Kraemer \& Jackson 1995; Beuther et al. 
2007), even including the rare isotopologue $^{15}$NH$_3$ (Mauersberger et 
al. 1986b; Schilke et al. 1991). 

A particularly outstanding hot core is associated with the star forming 
region W51-IRS\,2, which includes W51-North, W51d, W51d1, and W51d2 
(Zapata et al. 2009; their Fig.~1). At a distance of 6.1$\pm$1.3\,kpc 
(Imai et al. 2002) or 5.1$^{+2.9}_{-1,4}$\,kpc (Xu et al. 2009), 1\,pc 
subtends $\sim$40$''$.  With a rotational temperature $T_{\rm rot}$ $\sim$ 
300\,K, a source averaged column density of $N$(NH$_3$) $\sim$ 
10$^{19}$\,cm$^{-2}$, and a virial density of $n$(H$_2$) $\sim$ 
10$^{7.7}$\,cm$^{-3}$, W51-IRS2 is the source with the largest number of 
detected masing ammonia lines (Madden et al. 1986; Mauersberger et al. 1987; 
Wilson \& Henkel 1988). Here we focus on such non-thermally excited 
lines and report results from monitoring observations spanning a time 
interval of almost two decades and leading to the detection of a dozen 
new molecular maser lines.

\section{Observations}

The data were taken in November and December 1995, October 2008, 
August 2011, and in April 2012 with the 100-m Effelsberg 
telescope\footnote{Based on observations with the 100-m telescope of
the MPIfR (Max-Planck-Institut f{\"u}r Radioastronomie) at Effelsberg.} 
near Bonn/Germany. At the line frequencies observed (18.5 -- 31.4\,GHz), 
the beam size is 49$''$ -- 29$''$. The main beam brightness temperature scale 
was established by continuum cross scans toward 3C\,286 and NGC\,7027 
(flux densities were adopted from Ott et al. (1994), also accounting 
for a 0.5\%~yr$^{-1}$ secular decrease in the case of NGC\,7027), while 
1923+210 and 2145+06 were used as pointing sources. The pointing accuracy 
was better than 10$''$. For maser sources much more compact than the 
beam size, line shapes are not affected by pointing errors. In 1995 
and 1996, at frequencies above 26\,GHz, a single channel primary focus 
HEMT (Hot Electron Mobility Transistor) receiver was employed with a 
system equivalent temperature on a Jansky scale of $\sim $ 85\,Jy, 
including sky and ground radiation. Later, measurements at K-band 
(18--26\,GHz) were carried out with a dual channel (including both 
orthogonal linear polarizations) cooled primary focus HEMT receiver 
with a $T_{\rm sys}$ equivalent of $\sim$65\,Jy per channel.

\begin{table*}
\label{tab1}
\caption[]{Summary of NH$_3$ maser observations}
\begin{flushleft}
\begin{tabular}{cclcrccl}
\hline
Line      &  $\nu$    &  Epoch        & Channel        & $E_{\rm low}$/k & $S$   & $V_{\rm LSR}$  & $\Delta V_{1/2}$ \\
          &           &               & spacing        &                 &       &                &                  \\
($J,K$)   &  (GHz)    &               & (km\,s$^{-1}$) &    (K)          & (Jy)  & \multicolumn{2}{c}{km\,s$^{-1}$}  \\
\hline 
          &           &               &                &                 & \\
  (6,2)   & 18.884695 & 2012, Apr.4-8 &  0.97          &   578           & 0.22  & 54.39$\pm$0.02 & 1.26             \\
  (5,3)   & 21.285275 & 2012, Apr.4-8 &  0.86          &   380           & 0.30  & 57.01$\pm$0.02 & 1.37             \\
  (6,3)   & 19.757538 & 2012, Apr.4-8 &  0.93          &   551           & 0.35  & 56.50$\pm$0.01 & 1.44             \\
  (5,4)   & 22.653022 & 2012, Apr.4-8 &  0.81          &   342           & 0.52  & 57.00$\pm$0.02 & 1.54             \\
  (7,4)   & 19.218465 & 2012, Apr.4-8 &  0.95          &   713           & 0.09  & 54.74$\pm$0.05 & 1.31             \\
  (7,5)   & 20.804830 & 2012, Apr.4-8 &  0.88          &   664           & 3.10  & 54.44$\pm$0.02 & 0.95             \\
  (8,5)   & 18.808507 & 2012, Apr.4-8 &  0.97          &   892           & 0.14  & 54.38$\pm$0.02 & 1.44             \\
  (6,6)   & 25.056025 & 2008, Oct. 25 &  0.73          &   406           & 6.20  & 45.72$\pm$0.01 & 1.60             \\
          &           & 2012, Apr.4-8 &  0.73          &                 & 5.30  & 47.22$\pm$0.01 & 1.53             \\
  (7,6)   & 22.924940 & 2008, Oct. 25 &  0.80          &   605           & 1.95  & 46.02$\pm$0.01 & 1.77             \\
          &           & 2012, Apr.4-8 &  0.80          &   605           & 2.00  & 47.55$\pm$0.02 & 1.67             \\
  (8,6)   & 20.719221 & 2012, Apr.4-8 &  0.88          &   833           & 1.29  & 45.92$\pm$0.06 & 1.12             \\
  (9,6)   & 18.499390 & 2012, Apr. 5  &  0.99          &  1089           & 0.80  & 52.25$\pm$0.45 & 1.00             \\
          &           &               &                &                 & 6.80  & 54.37$\pm$0.04 & 1.19             \\
          &           &               &                &                 & 3.75  & 56.80$\pm$0.04 & 1.53             \\
          &           &               &                &                 & 1.15  & 61.23$\pm$0.20 & 3.70$\pm$0.50    \\
          &           & 2012, Apr. 7  &  0.99          &                 & 7.80  & 54.28$\pm$0.03 & 1.00             \\
          &           &               &                &                 & 4.90  & 56.59$\pm$0.06 & 1.02             \\
          &           &               &                &                 & 1.10  & 60.97$\pm$0.24 & 4.01$\pm$0.54    \\
  (7,7)   & 25.715182 & 2008, Oct. 23 &  0.71          &   535           & 1.70  & 45.60$\pm$0.02 & 1.46             \\
          &           &               &  0.02          &                 & 2.48  & 46.08$\pm$0.01 & 1.27$\pm$0.03    \\
          &           & 2008, Oct. 25 &  0.71          &                 & 2.27  & 45.66$\pm$0.05 & 1.57             \\
          &           & 2012, Apr.4-8 &  0.71          &                 & 0.20  & 47.29$\pm$0.11 & 1.38             \\
          &           &               &                &                 & 0.18  & 49.92$\pm$0.11 & 1.23             \\
  (9,7)   & 20.735452 & 2012, Apr.4-8 &  0.88          &  1020           & 0.08  & 54.72$\pm$0.08 & 0.88             \\
 (10,7)   & 18.285434 & 2012, Apr.4-8 &  1.00          &  1303           & 0.03  & 54.34$\pm$0.05 & 1.00             \\
  (9,8)   & 23.657471 & 2012, Apr.4-8 &  0.77          &   940           & 0.22  & 54.72$\pm$0.08 & 0.93             \\
  (9,9)   & 27.477943 & 1995, Nov. 28 &  0.53          &   848           & 0.16  & 43.76$\pm$0.07 & 2.14$\pm$0.15    \\
          &           &               &  0.13          &                 & 0.15  & 43.79$\pm$0.11 & 2.92$\pm$0.32    \\
          &           & 1995, Dec. 19 &  0.53          &                 & 0.12  & 43.63$\pm$0.19 & 2.70$\pm$0.41    \\
          &           & 2011, Aug. 31 &  0.94          &                 & 0.07  & 46.81$\pm$0.04 & 1.16             \\
  (10,9)  & 24.205287 & 2012, Apr.4-8 &  0.76          &  1133           & 0.34  & 46.06$\pm$0.04 & 1.20             \\
  (11,9)  & 21.070739 & 2012, Apr.4-8 &  0.87          &  1446           & 1.24  & 47.12$\pm$0.01 & 1.51             \\
 (12,12)  & 31.424943 & 1995, Nov. 28 &  0.46          &  1452           & 0.14  & 43.84$\pm$0.24 & 1.72$\pm$0.38    \\
          &           &               &  0.12          &                 & 0.12  & 44.30$\pm$0.30 & 2.35$\pm$0.71    \\
          &           & 1995, Dec. 19 &  0.47          &                 & 0.10  & 44.39$\pm$0.36 & 1.77$\pm$0.79    \\
          &           & 2011, Aug. 31 &  0.06          &                 & 0.60  & 46.87$\pm$0.02 & 0.81$\pm$0.03    \\
          &           &               &                &                 &       &                &                  \\
\hline
\end{tabular}
\end{flushleft}
The data are ordered with respect to $K$ from 2 to 12 and, 
within a $K$ ladder, with rising $J$ (see Sect.\,1). 
{\it Column 5}: $E_{\rm low}$/k: Energy above the ground state of the 
lower level of a given inversion doublet; k is the Boltzmann 
constant; ($E_{\rm up}$ -- $E_{\rm low}$)/k $\sim$ 1.0--1.5\,K. 
The beam size is $\theta_{\rm b}$ $\sim$ 50$''$ $\times$ 
(18\,GHz/$\nu$) with $\nu$ being the line frequency.
{\it Column 6}: Uncertainties in the flux densities are dominated by 
calibration errors, $\pm$ 15\% (1$\sigma$; see Sect.\,2) 
and are thus higher than the formal errors derived from Gaussian 
fits. Furthermore, if lines are spectrally unresolved, the 
given peak flux density is only a lower limit. 
{\it Column 7}: From Gaussian fits with formal standard deviations,
which are likely underestimating the real errors in case of 
spectrally unresolved features. {\it Column 8}: In many cases, the 
line width of the maser component is similar or smaller than the channel 
spacing displayed in column 4. In these cases, obtained line widths 
are upper limits, peak flux densities are lower limits, and no 
error is given. No maser emission was seen in the (1,1), (2,1),
(4,1), (5,1), (6,1), (2,2), (3,2), (4,2), (5,2), (3,3), (4,3), 
(7,3), (4,4), (6,4), (5,5), (6,5), (8,7), (10,8), (12,9) and (10,10))
lines down to 3$\sigma$ limits of order 0.05\,Jy in an individual 
0.7--1.0\,km\,s$^{-1}$ channel.
\end{table*}

For the early observations, the backend was a three level autocorrelator 
with 1024 spectral channels. Applied bandwidths ranged from 3.125 to 
25\,MHz. In 2008, and 2009, a Fast Fourier Transform Spectrometer (FFTS) was 
employed with 16384 channels for each polarization. Applied bandwidths 
were 50 and 500\,MHz. For the most recent data another FFTS was used, 
this time with a bandwidth of 2\,GHz and 32768 channels per polarization. 
This allowed us to cover the entire 18--26\,GHz range with a few 
frequency setups. Following weather conditions in a flexible way
(measurements are much less weather dependent at 18--20\,GHz 
than at higher frequencies), the frequency setup was changed quite often 
during the most recent observations with the 2\,GHz backend. These
covered parts of five consecutive days. Therefore, averaging data 
from several of these days, April 4--8, 2012, is mostly given as 
the epoch for these spectra. 

All data were taken in the position switching mode, employing scans 
of 6 minutes total duration and offset positions 10$'$ east or west. 
The calibration uncertainties are estimated to be $\pm$15\% (1$\sigma$).

\begin{figure}[t]
\vspace{0.0cm}
\centering
\resizebox{22.0cm}{!}{\rotatebox[origin=br]{-90}{\includegraphics{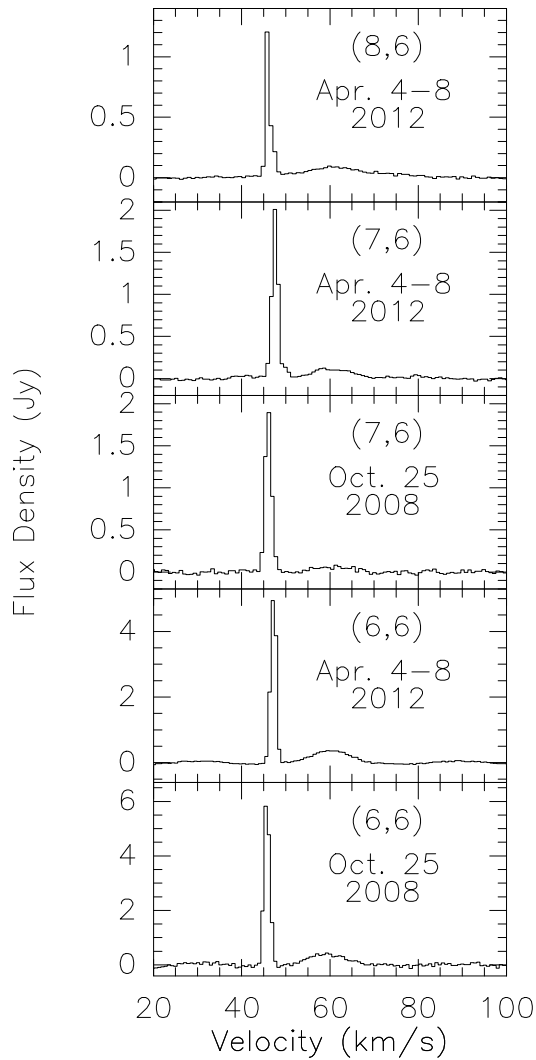}}}
\vspace{-1.0cm}
\caption{NH$_3$ spectra showing a new maser component near a Local Standard of 
Rest (LSR) velocity of 45\,km\,s$^{-1}$. Note the slight velocity shifts in the 
(6,6) and (7,6) emission as a function of time (see Sect.\,4.5). Less prominent 
quasi-thermal emission is centered near 60\,km\,s$^{-1}$. All spectra were taken 
at $\alpha_{\rm B1950}$ = 19$^{\rm h}$ 21$^{\rm m}$ 22\ffs19, $\delta_{\rm B1950}$ 
= +14$^{\circ}$ 25$^{\prime}$ 17\ffas0, corresponding to $\alpha_{\rm J2000}$ = 
19$^{\rm h}$ 23$^{\rm m}$ 39\ffs83, $\delta_{\rm J2000}$ = +14$^{\circ}$ 31$^{\prime}$ 
10\ffas1. For telescope beam sizes and channel spacings, see Table 1.
\label{fig1}}
\end{figure}

\section{Results}

\subsection{Are there lines caused by maser action?}

\begin{figure}[t]
\vspace{0.0cm}
\centering
\resizebox{22.0cm}{!}{\rotatebox[origin=br]{-90}{\includegraphics{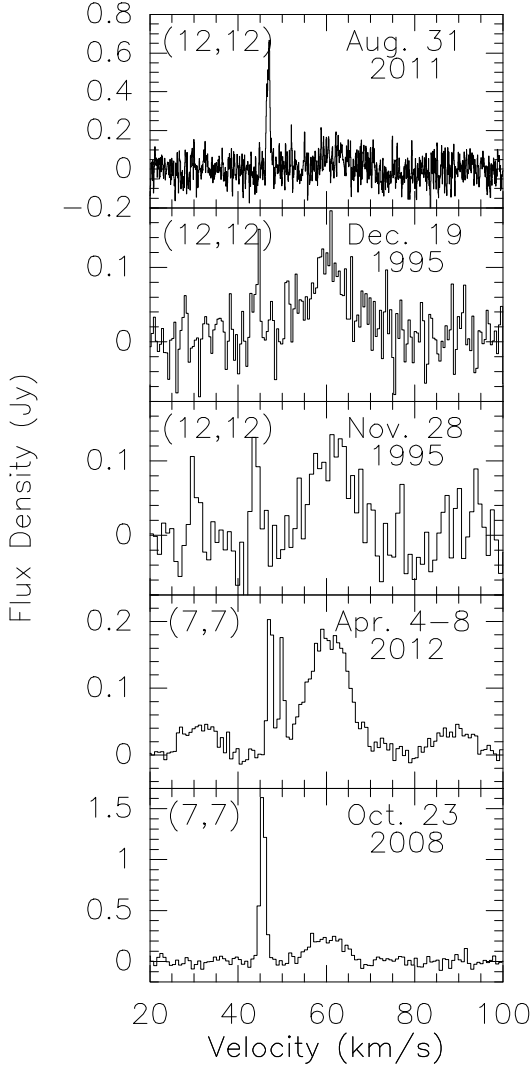}}}
\vspace{-0.7cm}
\caption{More NH$_3$ spectra showing the new 45\,km\,s$^{-1}$ 
component (see also the caption to Fig.~\ref{fig1}). In the second
lowest panel, the features near 30 and 90\,km\,s$^{-1}$ are 
quasi-thermally excited satellite lines of the (7,7) transition, 
caused by hyperfine splitting.  
\label{fig2}}
\end{figure}

Figures~\ref{fig1}--\ref{fig7} present spectra taken during a 
time interval of almost two decades, between 1995 and 2012. 
Shown is the complete set of maser transitions identified in 
the fully surveyed frequency interval between 18 and 26\,GHz
(with the two orthogonal polarizations being averaged) 
and by observations of the (9,9), (10,10), and (12,12) transitions 
between 27.5 and 31.4\,GHz. Lines, observing dates, channel 
spacings, and excitation above the ground state are summarized 
in Table~1. The characteristic rms noise level in a 
0.7--1.0\,km\,s$^{-1}$ channel is  of order 0.015\,Jy. Note that the 
(3,1) line at 22.23450\,GHz is blended by the much stronger 
6$_{1,6}$ $\rightarrow$ 5$_{2,3}$ line of H$_2$O. 

How can we be sure that each of the spectral lineshapes displayed 
is affected by population inversion? A strong argument favoring 
maser emission is a comparison of {\it all} detected lines. While 
Figs.~\ref{fig1}--\ref{fig7} display data from a total of 
19 NH$_3$ inversion lines, a significantly larger number
of such transitions was actually detected in the rich molecular
spectrum toward W51-IRS2 (see the caption to Table~1). All 
these spectra, e.g. the metastable ($J$=$K$) (1,1), (2,2), ... 
(5,5), and (10,10) transitions exhibit the well known comparatively 
broad $V_{\rm LSR}$ $\sim$ 60\,km\,s$^{-1}$ feature (e.g., 
Mauersberger et al. 1987). Since its intensity is closely related 
to excitation of the respective levels above the ground state 
(column 5 of Table~1 for the lines discussed in the following) 
and since the feature is not time variable within the limits of 
accuracy over the last quarter of a century, the emission 
must be quasi-thermal. 

Each of the 19 NH$_3$ transitions discussed in this paper shows
in addition to this quasi-thermal component at least one
additional spectral feature. These features tend to be narrower 
and almost all of those observed repeatedly show signs of variability
(see Sects.\,4.2 and 4.3). Furthermore, none of these 
additional components is seen in the inversion lines with 
lowest excitation (e.g., (1,1), (2,2), (3,3) or (5,5)), which 
were part of the frequency survey in 2012. All this strongly 
hints at maser emission, since such features, if quasi-thermal, 
would have to be seen in all inversion lines below a given 
excitation level. We thus adopt the maser classification 
for all 19 transitions displayed in Figs.~\ref{fig1}--\ref{fig7}.

\subsection{The spectra}

\begin{figure}[t]
\vspace{0.0cm}
\centering
\resizebox{22.0cm}{!}{\rotatebox[origin=br]{-90}{\includegraphics{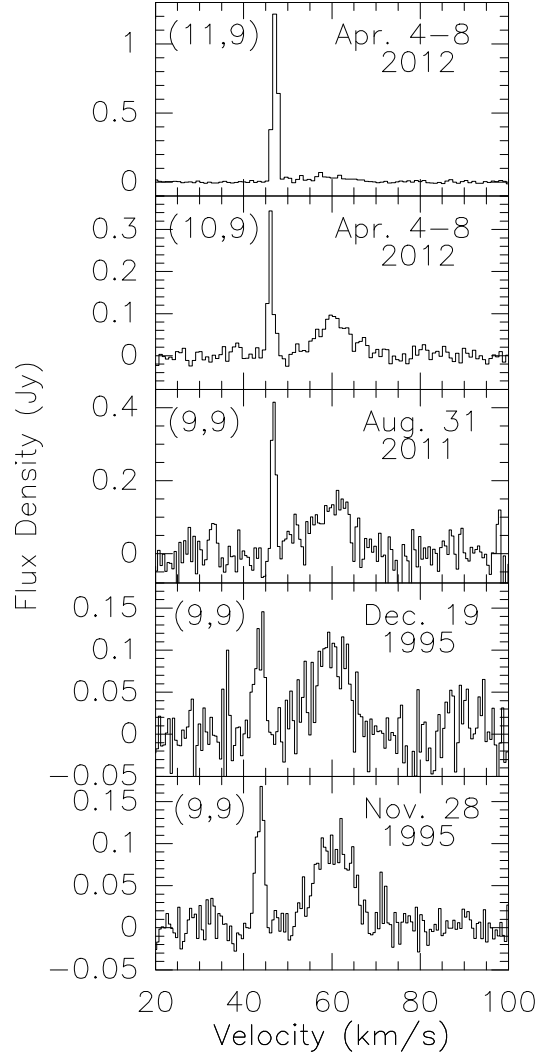}}}
\vspace{-0.7cm}
\caption{Additional NH$_3$ spectra showing the new 45\,km\,s$^{-1}$ 
component (see also Figs.~\ref{fig1} and \ref{fig2}). 
\label{fig3}}
\end{figure}

\begin{figure}[t]
\vspace{0.0cm}
\centering
\resizebox{9.2cm}{!}{\rotatebox[origin=br]{-90}{\includegraphics{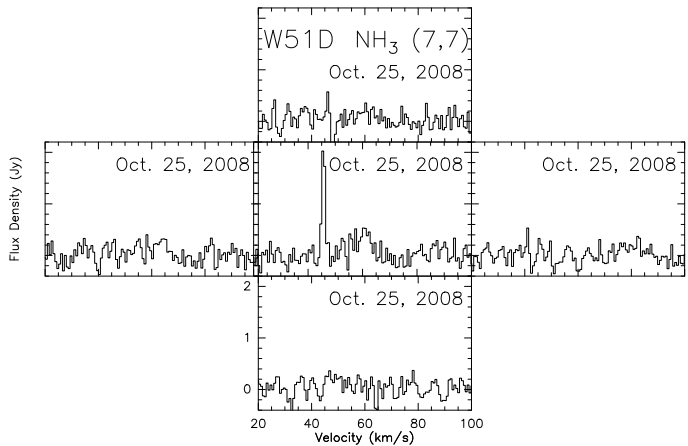}}}
\vspace{-0.3cm}
\caption{A map of NH$_3$ ($J$,$K$) = (7,7) emission with a spacing of 
30$''$ toward W51-IRS2. The velocity scale is Local Standard of Rest (LSR), 
the ordinate gives flux density (Jy). North is up and east is left.  
For beam size and channel spacing, see Table 1.
\label{fig4}}
\end{figure}

Among the 19 NH$_3$ maser lines identified by us in W51-IRS2,
13 have not been reported before from this source. The (6,3) 
and (9,6) lines (Madden et al. 1986), the (5,4), (7,5), and 
(9,8) lines (Mauersberger et al. 1987), and the (11,9) line 
(Wilson \& Henkel 1988) were known before this study. 
However, the non-thermal features in the (6,2), (5,3), (7,4), 
(8,5), (6,6), (7,6), (8,6), (7,7), (9,7), (10,7), (9,9), (10,9), 
and (12,12) transitions are new discoveries in W51-IRS2, with 
the (12,12) transition being the most highly excited NH$_3$ maser 
line so far detected. Among the newly identified maser lines 
in W51-IRS2 only the (6,6) and (8,6) transitions have an analog 
in NGC\,6334 (Beuther et al. 2007; Walsh et al. 2007) and 
possibly in IC\,342 (Lebr{\'o}n et al. 2011). All other lines 
have, to our knowledge, not yet been reported to show maser 
emission in outer space. Figs.~\ref{fig1}--\ref{fig7} thus reveal 
{\it eleven new molecular maser transitions}.

Figures~\ref{fig1}--\ref{fig3} display spectra, which contain 
a previously absent (e.g., Mauersberger et al. 1987) NH$_3$ velocity 
component in W51-IRS2, near a Local Standard of Rest velocity of 
$V_{\rm LSR}$ = 45\,km\,s$^{-1}$. In most cases the broad quasi-thermal 
feature near 60\,km\,s$^{-1}$ is dwarfed, sometimes to such an 
extent, that it is only barely visible (e.g., Fig.~\ref{fig1}). 
Figure~\ref{fig4} shows a five position map (spacing: 30$''$) of 
the ($J$,$K$) = (7,7) maser. Averaging the offset positions, the 
amplitude of the resulting signal relative to that from the central
position is 0.18$\pm$0.04. Using a Gaussian beam size of $\theta_{\rm b}$
= 35$''$, we obtain for a point source, in good agreement, 
a corresponding ratio of 0.14. Observations of the nearby source W51e 
in the (7,7) line on October 23, 2008, showed no 45\,km\,s$^{-1}$ maser 
component. The same holds for a (6,6) spectrum obtained on October 25, 
2008, toward W51e. The 5$\sigma$ limits are 0.27\,Jy and 0.88\,Jy in 
0.71\,km\,s$^{-1}$ and 0.73\,km\,s$^{-1}$ channels, respectively.

Figure~\ref{fig5} displays spectra of the 45\,km\,s$^{-1}$ component 
obtained with high spectral resolution (see also Sect.\,4.5). 
While the (12,12) spectrum from November 1995 has been smoothed, the 
original data, albeit being noisy, indicate that the line is as broad 
as the (9,9) feature measured at the same epoch. Figs.~\ref{fig6} and 
\ref{fig7} show spectra of ammonia lines with maser features being 
closer to the quasi-thermal ``systemic'' emission peaking near 
60\,km\,s$^{-1}$. Interestingly, the (3,3) and (10,8) lines, previously 
reported to exhibit maser emission at $\sim$54\,km\,s$^{-1}$
(Mauersberger et al. 1987; Zhang \& Ho 1995), now show exclusively 
the quasi-thermal 60\,km\,s$^{-1}$ component (Fig.~\ref{fig8}). While 
this requires a real change in the overall (10,8) line shape, the 
(3,3) maser was discovered through interferometric measurements and 
may be too weak and too close in frequency to the peak of the 
quasi-thermal emission to be detectable with single-dish telescopes.

A comparison of the two linearly polarized components of the masers 
measured between 18 and 26\,GHz shows no significant differences
in line shape in almost all cases. The exception is the (9,6) line,
where the 57\,km\,s$^{-1}$ component appears to be highly polarized
(see Fig.~\ref{fig9}). An alternative explanation, that there exist
two maser sources and that there is a beam squint between the two 
orthogonal polarisations, causing the different line shapes, can be 
excluded. Pointing and calibration measurements (Sect.\,2) do not
show any such position offset.

\section{Discussion}

\subsection{General considerations}

We have covered the entire 18--26\,GHz frequency band, which includes
all ammonia inversion doublets with low excitation. Nevertheless,
when employing a $\theta_{\rm b}$ $\sim$ 40$''$ beam, Table~1 
unambiguously tells us that only highly excited ammonia inversion 
doublets exhibit velocity components with inverted populations. 
Among the 19 detected maser lines, the ($J$,$K$) = (5,4) line at 
$E$ $\sim$ 340\,K is closest to the ground state.

\begin{figure}[t]
\vspace{0.0cm}
\centering
\resizebox{22.0cm}{!}{\rotatebox[origin=br]{-90}{\includegraphics{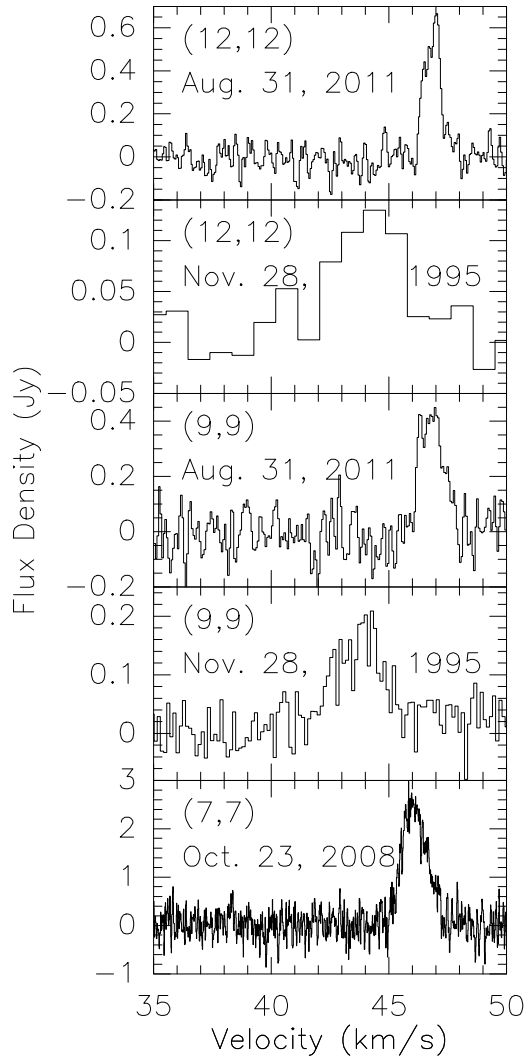}}}
\vspace{-1.0cm}
\caption{High velocity resolution spectra of NH$_3$ maser transitions showing the new
$\sim$45\,km\,s$^{-1}$ component. The upper and central profiles are extracted from 
Figs.~\ref{fig2} and \ref{fig3}, respectively. The second panel from the top shows 
a profile with eight contiguous channels smoothed (channel spacing: 0.96\,km\,s$^{-1}$,
otherwise, see Table 1). The original spectrum, albeit noisy, indicates a line 
width consistent with that shown in the spectrum of the (9,9) line measured 
on the same day. The Gaussian fit parameters of the (12,12) spectrum from November 28, 
1995, given in Table~1, were deduced from the original unsmoothed spectrum.
\label{fig5}}
\end{figure}

Another remarkable feature is the number of ortho- versus para-NH$_3$
maser lines. Even though para-NH$_3$ inversion doublets ($K$ = 1, 2, 4, 
5, 7, 8...) are twice as numerous as ortho-NH$_3$ doublets ($K$ = 3, 
6, 9...; the $K$ = 0 ladder is characterized by singlets), we get 
ten ortho- and only 9 para-NH$_3$ maser transitions. This may be 
understood in view of two important differences between the ortho- 
and para-species.  Firstly, the ortho levels have twice the statistical 
weight, thus providing larger column densities and possibly higher
maser amplification factors for ortho-to-para abundance ratios
of order unity (e.g., Umemoto et al. 1999; Goddi et al. 2011). 
Secondly, as already mentioned, the $K$=0 ladder contains single 
rotational states. In the gas phase, ortho- and para-NH$_3$ can 
exchange only very slowly (e.g., Cheung et al. 1969). The 
ortho-NH$_3$ $K$ = 0 single states therefore connect to one of the 
two levels in a $K$ = 3 inversion doublet, thus providing potentially 
significant deviations from quasi-thermally balanced populations at 
$K$ = 3 (e.g., Walmsley \& Ungerechts 1983). Within this context it 
is worth mentioning that the two strongest maser lines, those from 
the (6,6) and (9,6) inversion doublets, belong to the ortho-species.

Another noteworthy fact is that among the 19 maser lines detected 
by us, only four are metastable ($J$=$K$): three ortho-NH$_3$ [(6,6), 
(9,9), and (12,12)] transitions and one para-NH$_3$ [(7,7)] line (see 
Table~1). The vast majority of the maser lines detected 
in this study originates from non-metastable inversion doublets ($J$$>$$K$). 
Their populations decay rapidly to the metastable levels, so their excitation 
is complex and either requires extremely high densities ($\ga$10$^6$\,cm$^{-3}$)
and/or an intense infrared radiation field.

\begin{figure}[t]
\vspace{0.0cm}
\centering
\resizebox{22.0cm}{!}{\rotatebox[origin=br]{-90}{\includegraphics{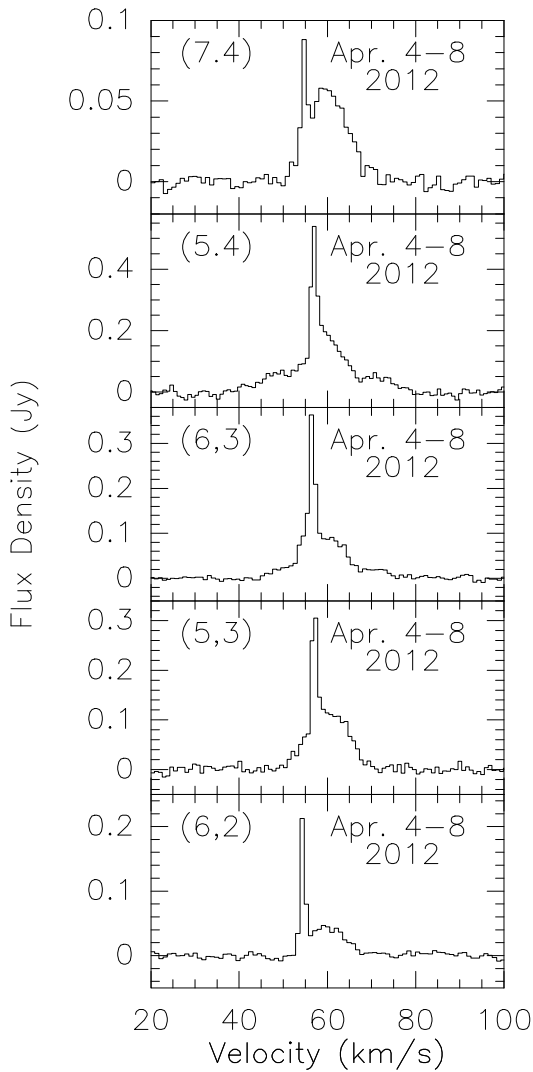}}}
\vspace{-0.7cm}
\caption{NH$_3$ maser spectra taken from W51-IRS2 not showing the 45\,km\,s$^{-1}$
maser component. For beam sizes and channel spacings, see Table 1.
\label{fig6}}
\end{figure}

\begin{figure}[t]
\vspace{0.0cm}
\centering
\resizebox{31.5cm}{!}{\rotatebox[origin=br]{-90}{\includegraphics{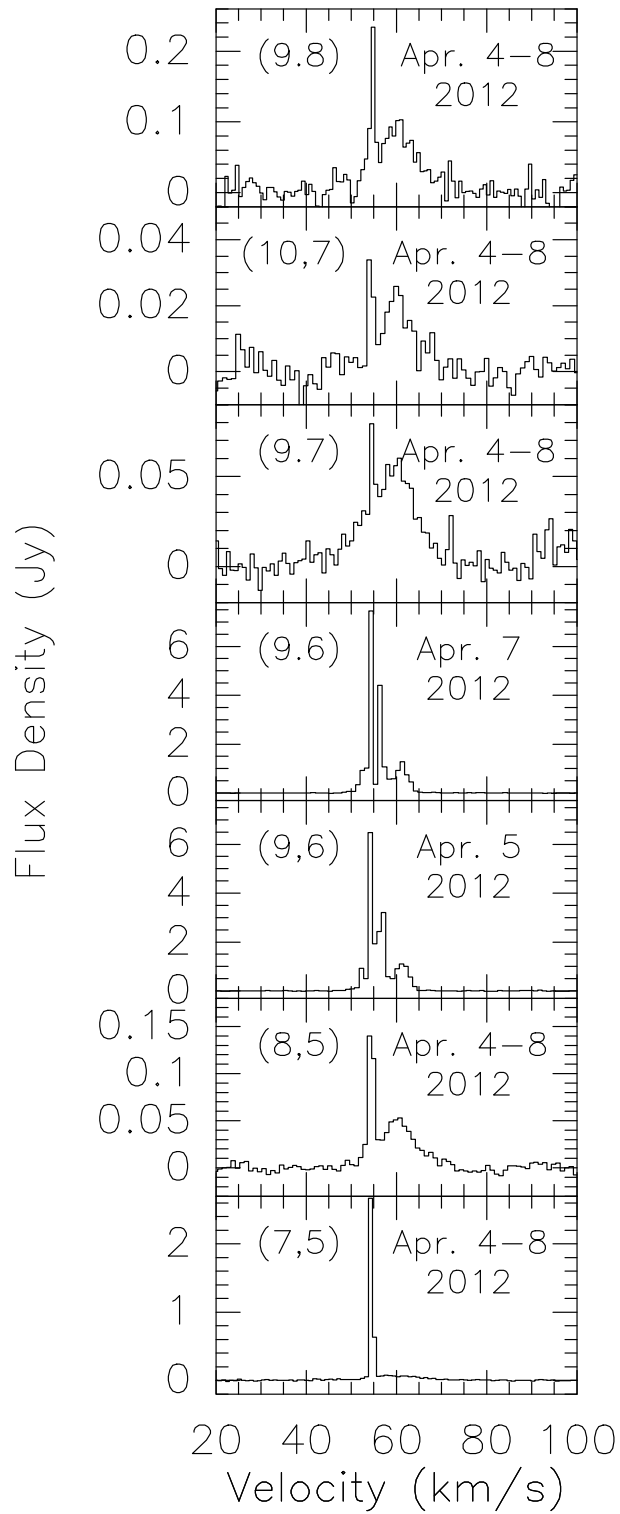}}}
\vspace{-1.3cm}
\caption{NH$_3$ maser spectra taken from W51-IRS2 not showing the 45\,km\,s$^{-1}$
maser component (see also Fig.~\ref{fig6}). 
\label{fig7}}
\end{figure}

\subsection{Variability on short time scales}

We seached for changes in the line profiles of our spectral
broadband survey during April 4--8, 2012. The (5,3), 
(5,4), (7,5), (7,6), (8,6), (9,7), and (11,9) transitions
remained stable, within the calibration uncertainties (Sect.\,2), 
over a time interval of four days. For the (6,3), (7,4), (6,6), 
and (7,7) lines the same holds for an interval of three days, 
while we see no significant change in the (6,2), (8,5), (9,8), 
and (10,9) transitions (and in the (7,7) transition in October
2008, see Figs.~\ref{fig2} and \ref{fig4}) after two days. 
However, the (9,6) line shows a significant variation in line 
shape within a time interval of only two days (Fig.~\ref{fig7}). 
To our knowledge this is the fastest variation so far seen 
in an ammonia maser and might indicate that the maser is 
due to exponential amplification of radiation, that is
unsaturated. Since it is one of the two strongest ammonia masers 
toward W51-IRS2, we could then speculate that all the 19 masers 
we have found are unsaturated. However, data with higher angular 
resolution are definitely needed to further clarify this point
(see, e.g., the arguments of Wilson et al. 1991 favoring a 
saturated (9,8) maser line).

We can assume that the time scale for intensity variations is equal
to the light travel time across the source. The main variation in the 
(9,6) line is seen near 55\,km\,s$^{-1}$. If we take a flux density 
of $S$ $\sim$ 2\,Jy (Fig.~\ref{fig7}), a beam size of 48\ffas6 (see 
caption to Table~1), a flux to main beam brightness temperature 
conversion factor of 1.5, and a distance of 5\,kpc (see Sect.\,1), 
we then obtain an angular scale of order 0\ffas1 and a  
corresponding brightness temperature of order 10$^6$\,K, similar to 
the limit obtained for the (9,8) line by Gaume et al. (1993).

\begin{figure}[t]
\vspace{-4.3cm}
\centering
\resizebox{27.5cm}{!}{\rotatebox[origin=br]{-90}{\includegraphics{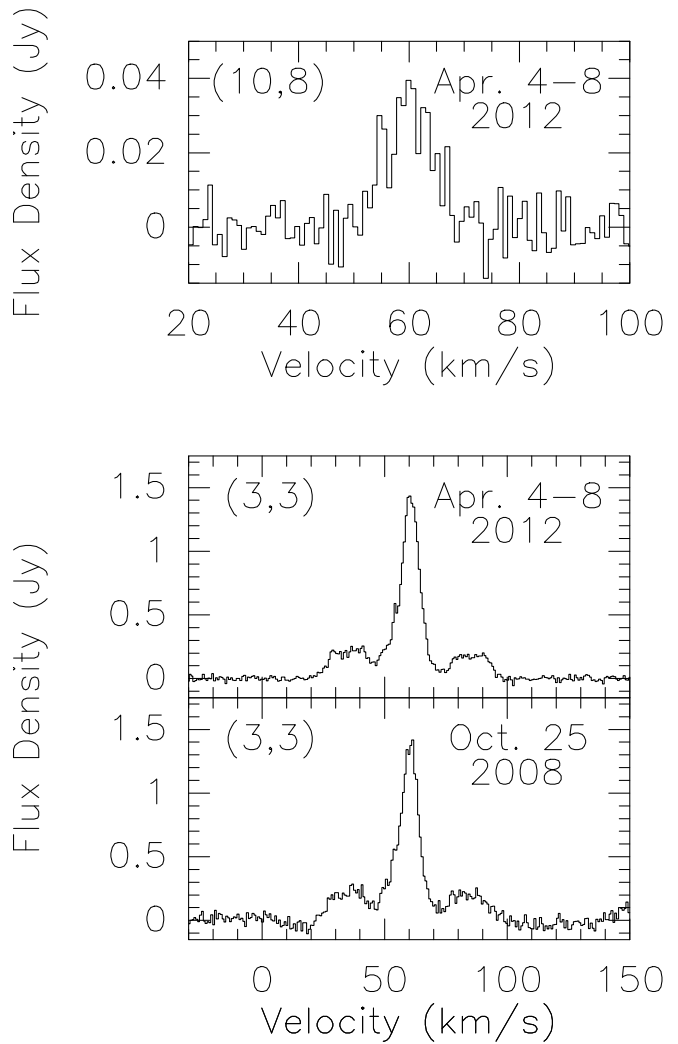}}}
\vspace{-0.7cm}
\caption{The two previously reported maser lines from W51-IRS2 
(Mauersberger et al. 1987; Zhang \& Ho 1995), which were
found to only exhibit quasi-thermal emission. The triple 
component line shape of the NH$_3$ (3,3) transition is caused 
by hyperfine splitting. Channel spacings are 0.88\,km\,s$^{-1}$
for the (10,8) and 0.77\,km\,s$^{-1}$ for the (3,3) profiles, 
respectively.
\label{fig8}}
\end{figure}

\subsection{Variability on long time scales}

\begin{figure}[t]
\vspace{-4.3cm}
\centering
\resizebox{27.5cm}{!}{\rotatebox[origin=br]{-90}{\includegraphics{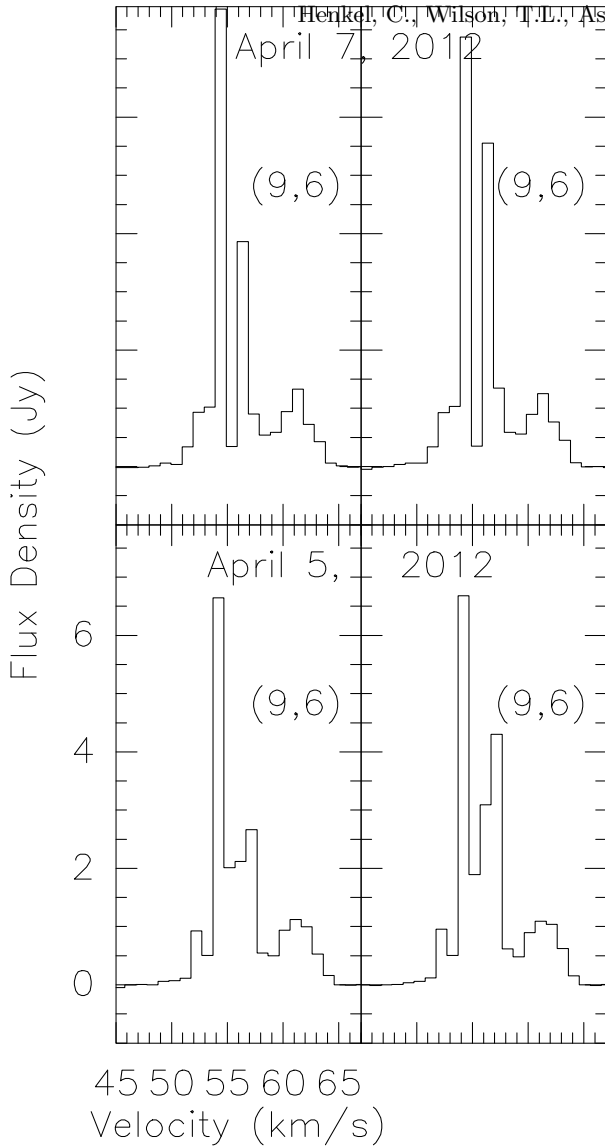}}}
\vspace{-0.7cm}
\caption{Linear polarisation in the 57\,km\,s$^{-1}$ component
of the (9,6) maser line on April 5 and 7, 2012. Shown are the 
two orthogonal components. The left panels display the horizontal 
(approximately east-west) and the right panels the vertical 
(approximately north-south) component, because the line was observed
close to transit. Since the line is strong, the difference is 
highly significant and is visible in each of the 20 (April 5) 
and 60 (April 7) individual subscans. For beam size and channel
spacing, see Table 1.
\label{fig9}}
\end{figure}

Beside the $V_{\rm LSR}$ $\sim$ 62\,km\,s$^{-1}$ component of the 
($J$,$K$) = (9,6) transition (Fig.~\ref{fig9}), which is too 
intense and time variable to be of quasi-thermal origin (cf. Madden
et al. 1985; Wilson et al. 1988), there are three additional 
families of maser lines. These show emission near 57\,km\,s$^{-1}$, 
near 54.5\,km\,s$^{-1}$, and near 45\,km\,s$^{-1}$ (see Table 1). In 
the following we discuss all components in order of decreasing 
velocity. 

High velocity features ($\ga$60\,km\,s$^{-1}$) were observed by 
Madden et al. (1986) and Wilson \& Henkel (1988) in W51-IRS2. The
Madden et al. (9,6) spectrum from late 1984 or early 1985 shows 
about half a dozen of narrow spikes between 50 and 62\,km\,s$^{-1}$. 
Flux densities are a few Jy. In this case, however, the angular 
resolution is 96$''$, so that the neighboring W51e region is also 
inside the beam (see, e.g., Fig.~1 of Gaume et al. 1993). In 
September 1987, Wilson \& Henkel (1988), using the Effelsberg 
telescope and therefore exclusively measuring the W51-IRS2 region, 
find a particularly strong component ($\sim$10\,Jy) near $V_{\rm LSR}$
$\sim$ 65\,km\,s$^{-1}$, which is, however, absent in July 1988.

$V_{\rm LSR}$ $\sim$ 57\,km\,s$^{-1}$ masers are seen in the (5,3), 
(6,3), (5,4), and (9,6) transitions, i.e. mostly in 
ortho-NH$_3$. There is rapid variability in the (9,6) line (see 
previous subsection). In the eighties (Madden et al. 1986; 
Wilson \& Henkel 1988), (6,3) emission was seen at 54\,km\,s$^{-1}$. 
The line has switched from one group of maser components
to another. The (5,4) feature has varied little during the past quarter 
of a century, while maser emission from the (11,9) line, detected
between August 1986 and October 1986 (Wilson \& Henkel 1988), is not
present in more recent spectra. 

The 54.5\,km\,s$^{-1}$ velocity component is found in eight maser
lines, namely in the (6,2), (7,4), (7,5), (8,5), (9,6), (9,7), 
(10,7), and (9,8) transitions. In earlier studies, also (3,3), 
(10,8), and (11,9) maser emission has been reported (Mauersberger 
et al. 1987; Wilson \& Henkel 1988; Zhang \& Ho 1995). The (6,3) line, 
which 25 years ago also contributed to the 54.5\,km\,s$^{-1}$ group 
of masers (Madden et al. 1986), is now seen at $\sim$57\,km\,s$^{-1}$
(see above). Except the (9,6) transition, all lines belong to the 
para-NH$_3$ species. Mauersberger et al. (1987) searched for (6,2) 
emission but the line remained undetected. Their upper limit, 
0.1\,Jy, is well below the 0.3\,Jy signal we have found. The 
(7,5) and (9,7) transitions did also not show a maser component 
(see Fig.~1 of Mauersberger et al. 1987). However, in July 1988, 
Wilson \& Henkel (1988) detected a 0.4\,Jy maser feature in the 
(7,5) line. Since then, the maser peak flux density has 
increased sixfold. The opposite trend, decreasing intensities 
with time, is seen in the (11,9), (9,8), and (10,8) lines.  
(11,9) maser emission was observed in 1986 (Wilson \& Henkel 
1988) but has faded. The (9,8) and (10,8) lines were much 
stronger $\sim$25 years ago. For the (9,8) line, already 
Wilson et al.  (1991) reported a moderate decrease by $\sim$30\% 
relative to measurements during the previous years. Now the decline 
has reached 90\% and the previously observed (10,8) maser is
now below our detection threshold.

Another eight transitions, the (6,6), (7,6), (8,6), (7,7), 
(9,9), (10,9), (11,9), and (12,12) lines, belong to the 
$V_{\rm LSR}$ $\sim$ 45\,km\,s$^{-1}$ group. Almost all of 
these are ortho-NH$_3$ transitions, again with one notable
exception, this time the (7,7) line. The 45\,km\,s$^{-1}$ 
component is the only one not seen in the (9,6) maser 
transition. For the broad quasi-thermal (8,6) feature near 
60\,km\,s$^{-1}$ Mauersberger et al. (1987) proposed 
anti-inversion, implying a low but still positive excitation
temperature and a high optical depth. It is thus remarkable that 
we find now, near 45\,km\,s$^{-1}$, strong (8,6) maser emission. 
In the (9,9) and (12,12) lines, which have been repeatedly 
measured, there is a tendency of rising peak flux densities 
in more recent observations. On the contrary, in spring 2012,
the (7,7) line has dramatically weakened, revealing a previously 
unseen second maser feature (Fig.~\ref{fig2}).

\subsection{Constraints for pumping scenarios}

We noted in Sect.\,4.2 that all observed masers in W51-IRS2 might 
be unsaturated. Wilson et al. (1991) proposed instead saturated
emission in the case of the (9,8) line. The two arguments were
a large line width ($\sim$0.4\,km\,s$^{-1}$; Wilson \& Henkel 1988)
and constant peak flux density. In the meantime, however, the line
has varied substantially and the line width is rather narrow. 
Measuring the line widths of all detected masers and comparing 
these, if possible, with previously taken data with high spectral
resolution will reveal the correlation between line width and 
peak flux density and will thus provide a definite answer to the 
question of maser saturation. This is part of a follow-up study.

\begin{figure}[t]
\vspace{0.0cm}
\centering
\resizebox{7.5cm}{!}{\rotatebox[origin=br]{0}{\includegraphics{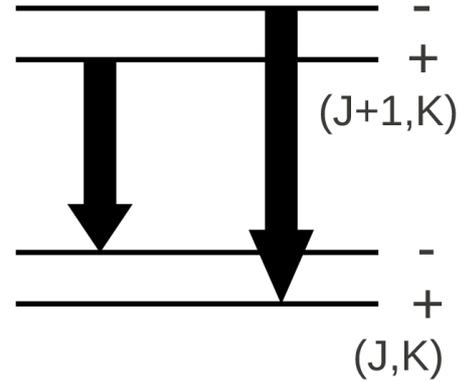}}}
\vspace{-0.0cm}
\caption{Selection rules for purely rotational transitions between two NH$_3$ 
inversion doublets belonging to the same K-ladder. Parities are given on the right 
hand side.
\label{fig10}}
\end{figure}

Figure~\ref{fig10} illustrates an important rule for radiative dipole
transitions within a K-ladder containing NH$_3$ inversion
doublets. The upper level of the upper doublet (the ($J$+1,$K$)
inversion doublet) is directly connected to the lower level 
of the lower doublet ($J$,$K$) and vice versa. Thus, if 
the upper doublet is inverted, the lower doublet should 
be anti-inverted. 

This is not confirmed by our measurements. The maser emission
at 57\,km\,s$^{-1}$ is observed in the (5,3) and (6,3) 
transitions. The 54.5\,km\,s$^{-1}$ component is seen 
in the (7,5) and (8,5) and in the (9,7) and (10,7) lines. 
In case of the 45\,km\,s$^{-1}$ component, we even find the 
(6,6), (7,6), (8,6) and the (9,9), (10,9), (11,9) transitions. 
It appears as if, contradicting the above mentioned rule, the
combination of ($J$+1,$K$) and ($J$,$K$) masers is not
forbidden but actually preferred. The former presence 
of the (9,8) and (10,8) maser lines and their simultaneous
weakening also points into this direction, even though truly
systematic variability studies of such maser pairs are still
not available.

As already noted by Madden et al. (1986), there are three
main pumping schemes which could, in principle, explain the 
observed NH$_3$ maser emission. These are (1) line overlap, 
(2) pumping by infrared radiation from the dust, and (3) 
pumping by collisions. Simulations of NH$_3$ maser 
emission encounter several problems. One is that collision rates 
are only known for inversion levels $J$ $\leq$ 6 (e.g., Danby 
et al. 1988). Another is the spectrum of the infrared radiation field.
For vibrational excitation of NH$_3$ by IR photons near $\lambda$
= 10$\mu$m (see Bally et al. 1987), the silicate dust 
feature may significantly perturb the radiation field, making 
any prediction highly speculative (Brown \& Cragg 1991). Madden 
et al. (1986) suggested for the NH$_3$ (10,6)$^+$--(9,6)$^-$ and 
(10,6)$^-$--(9,6)$^+$ transitions, the upper indices denoting 
parity, line overlaps. Such blends, if they are not common and 
only affect the (9,6) line, may explain the very particular 
behavior of this inversion line. Each maser family (Sect.\,4.3) 
is mostly composed of either ortho- or para-NH$_3$ transitions. 
This indicates that the two ammonia species are, as expected, 
reasonably well separated.

Schilke (1989) extrapolated NH$_3$ collision rates to high-$J$
levels and even included vibrational excitation (the $v_2$ = 1 
level), finding that vibrationally excited levels strongly 
affect the excitation of ground state inversion lines. For 
$T_{\rm dust}$ $\sim$ $T_{\rm kin}$, excitation temperatures 
can raise well above $T_{\rm dust}$ and $T_{\rm kin}$, in 
particular when temperatures above 160--180\,K are being reached. 
The latter does not hold for the bulk of the NH$_3$ bearing gas 
in Orion-KL (Hermsen et al. 1988) but is realized in W51-IRS2 
(Mauersberger et al. 1987). Interestingly, Schilke (1989) 
mentions the (9,8) line, which is easily inverted under such 
conditions, while the (10,8) line should not become a maser. 
This is consistent with our data from April 2012 (Figs.~\ref{fig7} 
and \ref{fig8}), but inconsistent with previous data also 
showing the (10,8) line to be inverted (Mauersberger et al. 1987).

Brown \& Cragg (1991) also studied the influence of vibrational
excitation on the ground state transitions. They could 
reproduce (6,3) maser emission and predicted quasi-thermal 
emission for the (4,3) and (5,3) lines. While the former is
consistent with our data, the latter is not (see Fig.~\ref{fig6}),
so that our data do not fully confirm their model in the case of 
W51-IRS2.

\begin{figure}[t]
\vspace{0.0cm}
\centering
\resizebox{7.5cm}{!}{\rotatebox[origin=br]{0}{\includegraphics{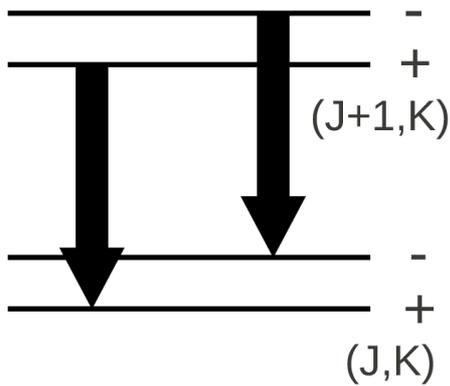}}}
\vspace{-0.0cm}
\caption{Pseudo-selection rules for rotational transitions between two NH$_3$ inversion 
doublets belonging to the same K-ladder, when vibrational excitation is involved. 
Parities are given on the right hand side. Excitation into a vibrationally excited 
level and subsequent decay leads to $\Delta$$J$ = 0, $\pm$1, and $\pm$2 pseudo-transitions
without parity change in the ground state. Shown are two R-branch transitions with 
$\Delta$$J$ = --1.
\label{fig11}}
\end{figure}

Following Schilke (1989), vibrational excitation leads to 
pseudo-selection rules in the ground state, where then $\Delta 
J$ = 0, $\pm$1, and $\pm$2 ``transitions'' are allowed without 
parity change. This directly contradicts the selection rules 
illustrated in Fig.~\ref{fig10} and leads to the situation in 
Fig.~\ref{fig11}, which predicts maser emission in several
adjacent $J$ levels within a K-ladder, as observed. We conclude 
that the measured series of masers in ($J$,$K$) and ($J$+1,$K$) 
lines, sometimes also including ($J$+2,$K$), are a direct 
consequence of vibrational excitation. Such a scenario was 
already proposed for the (3,3) and (4,3) lines of $^{15}$NH$_3$ 
in NGC~7538-IRS1 (Schilke et al. 1991). However, our case with 
several pairs and even triples of maser lines within a given 
K-ladder is a much more convincing one. The apparent contradiction
to Schilke with respect to the (10,8) line may be a result
of insufficient knowledge of collisional parameters (available
only up to the (6,6) line; Danby et al. 1988), spatial fine
structure, and the radiation field.

\subsection{NH$_3$ masers, the ambient molecular cloud, and a
feature with velocity drift}

Interferometric measurements to determine accurate positions 
have been carried out for the (3,3) and (9,8) maser lines 
(Gaume et al. 1993; Zhang \& Ho 1995). The (3,3) and (9,8) 
positions ($\alpha_{\rm J2000}$ $\sim$ 19$^{\rm h}$ 23$^{\rm m}$ 
39\ffas8, $\delta_{\rm J2000}$ $\sim$ +14$^{\circ}$ 31$^{\prime}$ 
05\ffas0) agree within the uncertainties of up to one arcsecond 
and may represent the position for the entire family of 
54.5\,km\,s$^{-1}$ maser lines (see Sect.\,4.3). These are 
associated with an ultracompact HII region (Wilson 
et al. 1991) that is also known to show 6.7\,GHz methanol 
emission (Phillips \& van Langevelde 2005). For the family
of 57\,km\,s$^{-1}$ masers an accurate position is not 
yet known. 

While frequently seen in late-type stars, W51-IRS2 is one 
of only three star forming regions known to show SiO maser emission
(Snyder \& Buhl 1974; Hasegawa et al. 1986). Such emission 
originates from vibrationally excited levels more than 1000\,K 
above the ground state and thus requires highly excited gas
(e.g., Goddi et al. 2009). There are no high NH$_3$ resolution 
data to prove a spatial correlation between $\sim$45\,km\,s$^{-1}$ 
NH$_3$ and SiO emission. However, there is a 45\,km\,s$^{-1}$ 
component seen in the SiO $v$=2 $J$ = 1$\rightarrow$0 maser line, 
possibly tracing an accelerating bipolar outflow close to 
the obscured protostar W51-North (Eisner et al. 2002; Zapata 
et al. 2009). The possible correlation between the highly excited 
45\,km\,s$^{-1}$ $v$ = 2 SiO maser component ($\alpha_{\rm J2000}$ 
= 19$^{\rm h}$ 23$^{\rm m}$ 40\ffas06, $\delta_{\rm J2000}$ = 
+14$^{\circ}$ 31$^{\prime}$ 05\ffas6) and NH$_3$ is supported
by the fact that the $\sim$45\,km\,s$^{-1}$ NH$_3$ masers are 
those with highest excitation, including the (12,12) transition at 
$E$/k $\sim$ 1450\,K (Table~1).

\begin{figure}[t]
\vspace{0.0cm}
\centering
\resizebox{15.0cm}{!}{\rotatebox[origin=br]{-90}{\includegraphics{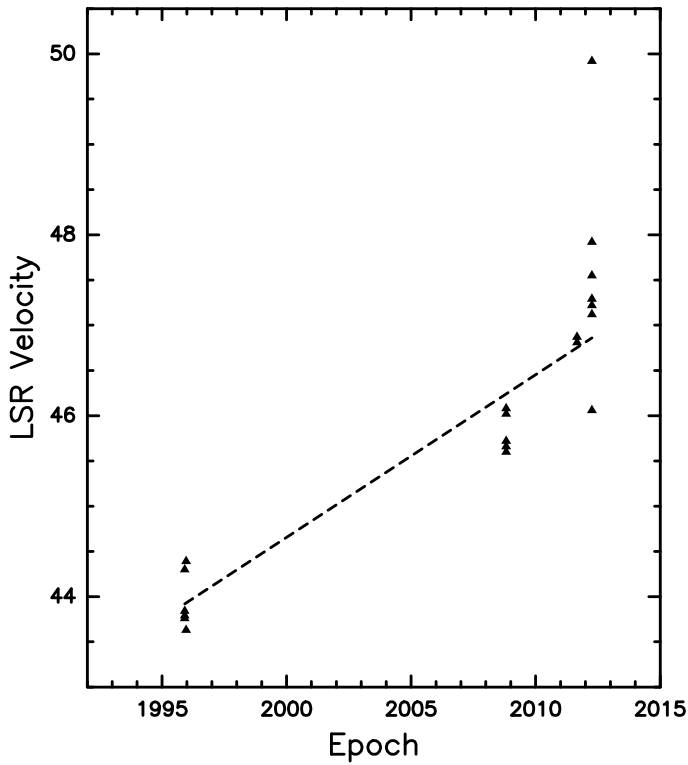}}}
\vspace{-0.0cm}
\caption{Radial velocity versus epoch for the $V_{\rm LSR}$ $\sim$ 45\,km\,s$^{-1}$
component with the data from Table~1. The dashed line provides an approximate
drift, based on the average velocity obtained in November 1995 and in April 2012.
For the latter epoch, the high velocity feature ($V_{\rm LSR}$ = 49.9\,km\,s$^{-1}$)
in the (7,7) line (Fig.~\ref{fig2}) was considered as an outlier and was not part 
of the average. 
\label{fig12}}
\end{figure}

Also related to this component, and perhaps the most striking 
aspect of our observations, is the velocity drift of the measured 
$\sim$45\,km\,s$^{-1}$ NH$_3$ features with time. This is seen in
Figs.~\ref{fig1}--\ref{fig3} and even more so in Fig.~\ref{fig5}. 
The drift is of order +0.2\,km\,s$^{-1}$\,yr$^{-1}$ (Fig.~\ref{fig12}). 
For July 1994, extrapolating the dashed line connecting data from 1995 
and 2012, we find a velocity of $V_{\rm LSR}$ $\sim$ 43.5\,km\,s$^{-1}$.
For July 1998, $V_{\rm LSR}$ $\sim$ 44.5\,km\,s$^{-1}$. Both 
values correspond to distinct SiO maser velocity components (Eisner 
et al. 2002), detected at these two epochs and might be related to
the lower velocity SiO feature monitored by Fuente et al. (1989; their 
Fig.~2), which shows a redward drift of order 0.4--0.5\,km\,s$^{-1}$\,yr$^{-1}$
between 1985 and 1989. In 1994, the NH$_3$ maser velocity is 
consistent with the strongest SiO maser feature observed by Eisner
et al. (2002), if we neglect the potential influence of missing flux 
in these high resolution data. All this supports a spatial 
correlation between SiO and the 45\,km\,s$^{-1}$ NH$_3$ maser 
component.

From SiO, Eisner et al. (2002) have identified a linear structure with 
a northern (more blue-shifted) and southern (more red-shifted) arm 
along a position angle of $\sim$105$^{\circ}$ with a velocity 
gradient of about 0.1\,km\,s$^{-1}$\,AU$^{-1}$ along each arm.
Assuming that the 45\,km\,s$^{-1}$ NH$_3$ feature is an outflow
region or a bullet belonging to this structure, the NH$_3$ velocity 
drift would correspond to a motion of 2\,AU\,yr$^{-1}$, roughly 
eastwards along P.A. $\sim$ 105$^{\circ}$. 

Eisner et al. (2002) interpreted the SiO distribution in terms of 
an accelerating outflow from a deeply obscured massive stellar 
source with a 3-dimensional acceleration of 
0.5\,km\,s$^{-1}$\,yr$^{-1}$ and 0.2\,km\,s$^{-1}$\,yr$^{-1}$
projected along the line-of-sight. The latter agrees remarkably well
with our ammonia data (Fig.~\ref{fig12}). The agreement is so 
good that we refrain from suggesting any other model as long
as future interferometric NH$_3$ measurements are not indicating a 
clearly different scenario.

Considering the proposed association of SiO and NH$_3$ masers,
we may ask whether such a phenomenon may be common. Toward
W51-IRS2, one out of several NH$_3$ maser velocity 
components appears to be related. Toward the other two
star forming SiO maser sources, Orion-KL and Sgr B2
(Hermsen et al. 1988; H{\"u}ttemeister et al. 1993; Goddi
et al. 2011), such NH$_3$ masers are not observed. The same 
holds for the circumstellar envelopes of late-type stars, 
which do not show NH$_3$ maser emission. This implies that 
we have detected a new phenomenon which might be rare. However, 
the number of known star forming regions exhibiting SiO maser 
emission is still too small to provide a sound statistical basis.

\subsection{Why does W51-IRS2 host so many ammonia maser sources?}

A question still to be addressed is why does W51-IRS2 host so 
many NH$_3$ masers? One potentially useful property may be a fortuitous
spatial configuration, with the densest part of the molecular hot core 
being located in front of an H{\sc ii} region (this may, however, 
not be the case for the 45\,km\,s$^{-1}$ component if it is 
associated with the SiO source (Eisner et al. 2002)). Another 
ingredient is likely the extremely high column density of NH$_3$ 
(Sect.\,1), which certainly facilitates the build up of sufficient
maser intensity. This holds in particular in the case of unsaturated maser 
emission (see Sects.\,4.2 and 4.4), where larger column densities 
lead to exponential effects. Possibly most important, however, is 
the kinetic temperature of the bulk of the ammonia emitting gas. With 
$T_{\rm kin}$ $\sim$ 300\,K (Mauersberger et al. 1987), it is much 
higher than, for example, in Orion-KL (Hermsen et al. 1988). This high 
temperature at densities, where dust and gas temperatures are likely
coupled, facilitates vibrational excitation by infrared photons near
10\,$\mu$m (see Sect.\,4.4). This may cause significant deviations 
from quasi-thermal conditions in the ground state (Schilke 1989; see 
also Mauersberger et al. 1988a and Schilke et al. 1990, 1992 for 
vibrationally excited NH$_3$ in the cooler Orion-KL hot core). Line 
overlaps may also play a role. Note that the excitation above the ground 
state of the NH$_3$ (12,12) maser line (Figs.~\ref{fig2} and \ref{fig5}) 
is of the same order of magnitude as that of the first vibrationally 
excited state.

\section{Conclusions}

Using the 100-m telescope at Effelsberg, we have searched for NH$_3$
maser emission in the prominent massive star forming region W51-IRS2.
The main results from this study, covering 16 years of monitoring, 
are briefly mentioned: 

\begin{itemize}

\item A total of 19 NH$_3$ maser lines was detected. Eleven of these
inversion transitions have been identified as masers for the first 
time in the interstellar medium. 

\item The (9,6) line has been found to vary significantly during
a time interval of only two days. This is the fastest change so 
far detected in an ammonia maser. Furthermore, its 57\,km\,s$^{-1}$
velocity component shows strong linear polarisation, which is not
found in any of the other lines. While variations in other transitions 
are slower, not only ortho- but also para-NH$_3$ transitions show
drastic changes during the last 16 years. The (12,12) line,
at $\sim$1450\,K above the ground state, is the most highly excited
ammonia maser line so far detected. 

\item Apart from the (9,6) line, which is unique among 
the 19 transitions reported here, there are three
families of maser transitions, those exhibiting emission at 
57, 54.5 and 45\,km\,s$^{-1}$. The latter family was not known
previously.

\item There are many maser pairs in the same $K$ ladder,
such as ($J$,$K$) and ($J$+1,$K$). In some cases, this extends
to a triple, including ($J$+2,$K$). This observation gives strong
support for vibrational excitation, induced by $\lambda$ 
$\sim$ 10$\mu$m photons, as the main pumping mechanism. 

\item The 45\,km\,s$^{-1}$ component masers show a secular drift
of about 0.2\,km\,s$^{-1}$\,yr$^{-1}$ and may arise close to 
the SiO maser source. Possible scenarios explaining the drift
in terms of outflowing gas or orbital motion are outlined.

\item The most likely reason that W51-IRS2 hosts many more NH$_3$
masers than any other known source is the combination
of a very high temperature ($\sim$300\,K, ideal for 10$\mu$m
photons needed to vibrationally excite NH$_3$), an exceptionally 
high NH$_3$ column density, and possibly also a fortuitous location
with respect to continuum sources such as hyper-compact 
H{\sc ii} regions.

\end{itemize}

Besides measuring accurate line widths of many of the newly
discovered maser lines to discriminate between saturated and
unsaturated maser emission, the most important follow-up 
will be high angular resolution measurements. Only these will
allow us to determine precise positions, which are essential
to interprete the data in a broader context, also including 
other astrophysical tracers, and to explain the origin of the 
enigmatic velocity drift.

\acknowledgements
We wish to thank A. Kraus for helpful comments related to the 
data taken at the Effelsberg 100-m telescope. We are also 
happy to thank C. Goddi and C.M. Walmsley for carefully 
reading the manuscript and providing excellent comments.


\begin{thebibliography}{}
 \bibitem[1987]{xyz}
  Bally, J., Arens, J. F., Ball, R., Becker, R., \& Lacy, J. 1987,
  ApJ, 323, L73
 \bibitem[2007]{xyz}
  Beuther, H., Walsh, A.J., Thorwirth, S., et al. 2007, A\&A, 466, 989
 \bibitem[1999]{xyz}
  Brown, P. D., Charnley, S. B., \& Millar, T.J. 1988, MNRAS, 231, 409 
 \bibitem[1991]{xyz}
  Brown, R. D., \& Cragg, D. M. 1991, ApJ, 378, 445
 \bibitem[1992]{xyz}
  Cesaroni, R., Walmsley, C. M., \& Churchwell, E. 1992, A\&A, 256, 618
 \bibitem[1969]{xyz}
  Cheung, A. C., Rank, D. M., Townes, C. H., Knowles, S. H., \& Sullivan, W. T.
  1969, ApJ, 157, L13
 \bibitem[1988]{xyz}
  Danby, G., Flower, D.R., Valiron, P., Schilke, P., \& Walmsley, C.M. 1988, 
  MNRAS, 235, 229 
 \bibitem[2002]{xyz}
  Eisner, J. A., Greenhill, L. J., Herrnstein, J. R., Moran, J. M., \& Menten, K. M. 
  2002, ApJ, 569, 334
 \bibitem[1989]{xyz}
  Fuente, A., Mart\'{\i}n-Pintado, J., Alcolea, J., \& Barcia, A. 1989, A\&A, 223, 321
 \bibitem[1993]{xyz}
  Gaume, R. A., Johnston, K. J., \& Wilson, T. L. 1993, ApJ, 417, 645
 \bibitem[2009]{xyz}
  Goddi, C., Greenhill, L.J., Chandler, C. J., et al. 2009, ApJ 698, 1165
 \bibitem[2011]{xyz}
  Goddi, C., Greenhill, L.J., Humphreys, E. M. L., Chandler, C. J., \& Matthews 
  2011, ApJ 739, L13
 \bibitem[1999]{xyz}
  Guilloteau, S., Wilson, T. L., Martin, R. N., Batrla, W. \& Pauls, T. A., 1983,
  A\&A, 124, 322
 \bibitem[1005]{xyz}
  Greenhill, L. J., Henkel, C., Becker, R., Wilson, T. L., \& Wouterloot, J. G. A. 1995,
  A\&A, 304, 21
 \bibitem[1984]{xyz}
  Haschick, A. D., Baan, W. A., \& Peng, E. W. 1994, ApJ, 437, L35
 \bibitem[1986]{xyz}
  Hasegawa, T., Morita, K., Okumura, S., et al. 1986, in {\it Masers, Molecules, and 
  Mass Outflows in Star Forming Regions}, ed. A.D. Haschick (Westford, Haystack), 275
 \bibitem[1999]{xyz}
  Henkel, C., Mauersberger, R., Wilson, T. L., et al. 1987a, A\&A, 182, 299
 \bibitem[1999]{xyz}
  Henkel, C., Wilson, T. L., \& Mauersberger, R. 1987b, A\&A, 182, 137
 \bibitem[1999]{xyz}
  Hermsen, W., Wilson, T. L., Walmsley, C. M., \& Henkel, C. 1988, A\&A, 201, 285
 \bibitem[1983]{xyz}
  Ho, P. T. P., \& Townes, C.H. 1983, ARA\&A, 21, 239
 \bibitem[1994]{xyz} 
  Hofner, P., Kurtz, S., Churchwell, E., Walmsley, C.M., \& Cesaroni, R. 1994, A\&A, 429, L85
 \bibitem[1999]{xyz}
  H{\"u}ttemeister, S., Wilson, T. L., Henkel, C., \& Mauersberger, R. 1993, 
  A\&A, 276, 445
 \bibitem[1999]{xyz}
  H{\"u}ttemeister, S., Wilson, T. L., Mauersberger, R., et al. 1995, A\&A,
  294, 667
 \bibitem[2002]{xyz}
  Imai, H., Watanabe, T., Omodaka, T., et al. 2002, PASJ, 54 741 
 \bibitem[1995]{xyz}
  Kraemer, K.E., Jackson, J.M. 1995, ApJ, 439, L9
 \bibitem[2011]{xyz}
  Lebr{\'o}n, M., Mangum, J. G., Mauersberger, R., et al. 2011, A\&A, 534, A56
 \bibitem[1999]{xyz}
  Madden, S. C., Irvine, W. A., Matthews, H. E., Brown, R. D., \& Godfrey, P. D.
  1986, ApJ, 300, L79
 \bibitem[1986]{xyz}
  Mauersberger, R., Henkel, C., Wilson, T. L., \& Walmsley, C. M. 1986a, A\&A, 162, 199
 \bibitem[1999]{xyz}
  Mauersberger, R., Wilson, T. L., \& Henkel, C. 1986b, A\&A, 160, L13
 \bibitem[1999]{xyz}
  Mauersberger, R., Henkel, C., \& Wilson, T. L. 1987, A\&A, 173, 352
 \bibitem[1999]{xyz}
  Mauersberger, R., Henkel, C., \& Wilson, T.L. 1988a, A\&A, 205, 235
 \bibitem[1999]{xyz}
  Mauersberger, R., Wilson, T. L., \& Henkel, C. 1988b, A\&A, 201, 123
 \bibitem[1995]{xyz}
  Miyoshi, M., Moran, J., Herrnstein, J., et al. 1995, Nature, 373, 127
 \bibitem[1999]{xyz}
  Ott, M., Witzel, A., Quirrenbach, A., et al. 1984, A\&A, 284, 331
 \bibitem[2005]{xyz}
  Phillips, C., van Langevelde, H. J. 2005, Ap\&SS, 295, 225
 \bibitem[1989]{xyz}
  Schilke, P. 1989, Diploma Thesis, {\it Ammoniak in warmen Molek{\"u}lwolken},
  Univ. of Bonn/Germany
 \bibitem[1990]{xyz}
  Schilke, P., Walmsley, C. M., Wilson, T. L., Mauersberger, R. 1990, A\&A,
  227, 220
 \bibitem[1999]{xyz}
  Schilke, P., Walmsley, C. M., \& Mauersberger, R. 1991, A\&A, 247, 516
 \bibitem[1992]{xyz}
  Schilke, P., G{\"u}sten, R., Schulz, A., Serabyn, E., \& Walmsley, C.M.
  1992, A\&A, 261, L5
 \bibitem[1974]{xyz}
  Snyder, L. E., \& Buhl, D. 1974, ApJ, 189, L31
 \bibitem[1999]{xyz}
  Umemoto, T., Mikami, H., Yamamoto, S., \& Hirano, N. 1999, ApJ, 525, L105
 \bibitem[1983]{xyz}
  Walmsley, C. M., \& Ungerechts, H. 1983, A\&A, 122, 164
 \bibitem[1999]{xyz}
  Walmsley, C.M., Hermsen, W., Henkel, C., et al. 1987, A\&A, 172, 311
 \bibitem[2007]{xyz}
  Walsh, A. J., Longmore, S. N., Thorwirth, S., Urquhart, J. S., \& 
  Purcell, C. R. 2007, MNRAS, 382, L35
 \bibitem[1999]{xyz}
  Wilson, T. L., \& Henkel, C. 1988, A\&A, 206, L26
 \bibitem[1999]{xyz}
  Wilson, T. L., Batrla, W., \& Pauls, T. A. 1982, A\&A, 110, L20
 \bibitem[1991]{xyz}
  Wilson, T. L., Gaume, R., \& Johnston, K. J. 1991, A\&A, 251, L7
 \bibitem[1999]{xyz}
  Wilson, T. L., Henkel, C., H{\"u}ttemeister, S., et al. 1993, A\&A, 276, L29
 \bibitem[2000]{xyz}
  Wilson, T. L., Gaume, R. A., Gensheimer, P. \& Johnston, K. J. 2000, A\&A, 251, L7
 \bibitem[2006]{xyz}
  Wilson, T. L., Henkel, C., \& H{\"u}ttemeister, S. 2006, A\&A, 460, 533
 \bibitem[1996]{xyz} 
  Wilson, T. L., Rohlfs, K., \& H{\"u}ttemeister 2008, {\it Tools of Radio 
  Astronomy}, 5th Edition, Springer, p404  
 \bibitem[2009]{xyz}
  Xu, Y., Reid, M. J., Menten, K. M., et al. 2009, A\&A, 693, 413
 \bibitem[2009]{xyz}
  Zapata, L.A., Ho, P. T. P., Schilke, P., et al. 2009, ApJ 698, 1422
 \bibitem[1995]{xyz}
  Zhang, Q., \& Ho, P. T. P. 1995, ApJ, 450, L63
 \bibitem[1997]{xyz}
  Zhang, Q., \& Ho, P. T. P. 1997, ApJ, 488, 241

\end{thebibliography}
\end{document}